\documentclass[vecphys]{svmult}

\usepackage{makeidx}         
\usepackage{graphicx}        
\usepackage{multicol}        
\usepackage[bottom]{footmisc}

\makeindex             

\def\frac#1#2{{\textstyle{{#1}\over {#2}}}}
\def\goto{\rightarrow}

\def\lsim{\mathrel{\rlap{\lower4pt\hbox{\hskip1pt$\sim$}}
    \raise1pt\hbox{$<$}}}
\def\gsim{\mathrel{\rlap{\lower4pt\hbox{\hskip1pt$\sim$}}
    \raise1pt\hbox{$>$}}}
\def\sqr#1#2{{\vcenter{\vbox{\hrule height.#2pt
         \hbox{\vrule width.#2pt height#1pt \kern#1pt
         \vrule width.#2pt}
         \hrule height.#2pt}}}}

\def\beq{\begin{equation}}
\def\eeq{\end{equation}}
\def\beqa{\begin{eqnarray}} 
\def\eeqa{\end{eqnarray}}

\def\laq{\raise 0.4 ex \hbox{$<$}\kern -0.8 em\lower 0.62 ex\hbox{$\sim$}}
\def\gaq{\raise 0.4 ex \hbox{$>$}\kern -0.7 em\lower 0.62 ex\hbox{$\sim$}}

\begin{document}

\title*{The mystical formula and the mystery of Khronos}
\author{Orfeu Bertolami}
\institute{Instituto Superior T\'ecnico, Departamento de F\'\i sica, \\
Av. Rovisco Pais 1, 1049-001, Lisboa, Portugal \\
\texttt{orfeu@cosmos.ist.utl.pt}}


\maketitle

\section*{Abstract}

In 1908, Minkowski put forward the idea that 
invariance under what we call today 
the Lorentz group, $GL(1,3, {\bf R})$, would be more meaningful in a four-dimensional 
space-time continuum. This suggestion implies that space and time are intertwined entities  
so that, kinematic and dynamical quantities can be expressed as vectors, 
or more generally by tensors, in the four-dimensional space-time. Minkowski also 
showed how causality should be structured in the four-dimensional vector space.  
The mathematical formulation proposed by Minkowski made its generalization to curved spaces 
quite natural, leaving the doors to the General Theory of Relativity and many other developments ajar.

Nevertheless, it is remarkable that this deceptively simple formulation 
eluded many researchers of space and time, and goes against our every day experience and perception, 
according to which space and time are distinct entities. In this contribution, we discuss these 
contradictory views, analyze how they are seen in contemporary physics and comment on the challenges that 
space-time explorers face. 
  


\section{The mystical formula}
\label{sec:1}

On the 21st of September 1908, at his address to the 80th Assembly of German Natural 
Scientists and 
Physicians in Cologne, Hermann Minkowski (1864 - 1909) presented his World Postulate, 
{\it Weltpostulat}, according to which the invariance under what we call today 
the Lorentz group $GL(1,3, {\bf R})$, 
would be more meaningful in a 4-dimensional 
space-time continuum \cite{Minkowski}. This follows the very spirit of special relativity, in 
that the independence of the laws of physics on the 
velocity of inertial frames requires that space and time are indissoluble 
concepts and are related to each other by the maximum attainable 
particle velocity. In his own words:

``The views of space and time which I wish to lay down for you have sprung 
from the soil of experimental physics, and therein lies their strength. 
They are radical. Henceforth space by itself, and time by itself, are 
doomed to fade away into mere shadows, and only a kind of union of the 
two will preserve an independent reality.''

To further stress the somewhat unusual nature of his proposal, Minkowski simply states that 
``no one can ever refer to a {\it  Space} without its {\it Time} or to a {\it Time} without its {\it Space}'' 
and actually, much more emphatically, ``the essence 
of the World Postulate, which is pregnant with mathematical implications, could be dressed in the mystical
formula:'' 
\beq
3 \times 10^5~km = \sqrt{-1}~sec. ~,
\eeq 
where the imaginary unit arises as Minkowski chooses to view space-time as strictly Euclidean, in its signature 
and in its lack of curvature. Thus, an event in the space-time continuum should be referred to as a world-point, ``Welt-punkt'', 
while its evolution in space-time continuum through a world-line, ``Welt-linie''. Thus, according to 
Minkowski, ``all the world presents itself quite explicitly through world-lines'' to the point that in his opinion,  
``physical laws would find their most comprehensive formulation through the reciprocal relationships of world-lines''. 

It is clear that the suggestion of a space-time continuum 
represents a further unification of concepts in physics (for a discussion on the inconsistency of a 3-dimensional world see Ref. 
\cite{Petkov}). 
Indeed, special relativity allowed for a unified description of the laws of physics as 
well as for an unique formulation of mass, energy and momenta, thanks to the invariance of the maximum 
attainable particle velocity, $c_{ST}$, the Relativity Principle. It is important to remember that from the 
Relativity Principle, in any physical setting, distances 
can be measured by clocks and mirrors. Furthermore, an immediate implication of 
this order of ideas is that if space is isotropic, then any attempt to 
measure the time difference in the time of travel of light of equal distance paths 
would yield, irrespective of the direction, a null result. The most recent 
experimental 
attempts to measure deviations from this null outcome have shown that it holds up to 
a few parts in $10^{-9}$ \cite{Peters}. 
Indirect experiments involving, for instance, ultra high-energy cosmic rays yield 
even more impressive limits, 
actually $1.7 \times 10^{-25}$ (see e.g. Ref. \cite{BCarvalho00} and references therein). 

At this point, two comments are in order. The first one refers to the fact that the identification of the maximum 
attainable particle velocity with the speed of light, $c$, is only possible because, 
up to the current experimental precision, the photon mass vanishes and electromagnetism is an exact 
abelian gauge theory. If this were not the case, these two velocities could not be the same. Of course, historically, 
these two velocities were not initially distinguished and current bounds on the photon mass   
are compatible with this identification (see for instance \cite{BMota}). Naturally, the same can be said about the identification 
of the speed of light and the velocity of propagation of gravitational waves in vacuum \cite{EllisUzan}. Thus, given the 
present bounds on the the photon (and also the graviton) mass we shall simply identify $c_{ST}$ with $c$.  

The second comment refers to the fact that the Standard Model (SM) vacuum being non-trivial, might not 
respect Lorentz invariance and hence correspond to a sort of preferred frame \cite{Consoli}. Alternatively,  
one can consider that only particle Lorentz invariance is physically meaningful, a perspective which 
allows for an extension of the SM compatible with the 
spontaneous breaking of Lorentz invariance, without implying the existence of a 
preferred frame \cite{CollKost}. Of course, this possibility would lead to distinct experimental signatures, 
which so far have not been observed (see e.g. Refs. \cite{Kost07} for comprehensive discussions). 

So according to Minkowski, the motion of particles in the space-time continuum correspond to lines, 
{\it world lines}, from a given point in space-time where the original {\it event} took place. Past and future and 
hence causality are referred to this original event. In terms of the metric 
\begin{equation}
\label{Minkowski} ds^2 = -c^2dt^2 + dx^2 + dy^2 + dz^2 ~,
\end{equation}
space-time admits {\it light-like} world lines, for which $ds^2=0$, {\it time-like} interval, for which $ds^2 <0$, 
and a {\it space-like} interval, for which $ds^2 >0$. Thus, in space-time diagrams, where time is depicted in the vertical axis and 
space in the horizontal one, light travels in the cone, the {\it light-cone} or null curves described, for a given time 
interval $\Delta t$, by 
$c\Delta t= \pm \sqrt{\Delta x^2 + \Delta y^2 + \Delta z^2}$. Events within this cone are time-like and given that observers move with 
a relative velocity $v <c$, word-lines within this cone connect events in the past or in the future of each other, whether they 
precede or succeed each other. An event, say $A$, outside the light-cone cannot influence or be influenced by any other event 
separated by $A$ by a space-like interval. 
 
Clearly, these relationships have an absolute and global nature 
given the independence of the velocity of light on the velocity of the frame of 
reference; however, special relativity makes the concept of an absolute simultaneity impossible and thus the idea of an universal 
present. Of course, as already mentioned, this together with the fact that time flows at different rates for different 
observers and that likewise, the perception of space is also tied up with the relative motion of different observers, 
drives one away from 
Newton's (1643 - 1727) conception of absolute time, defined in the first book of his 
{\it Principia Mathematica } in 1687: ``Absolute, true, and mathematical time, in and of itself 
and of its own nature, without reference to anything external, flows uniformly and by another name is called duration''.         

The insightful formulation of Minkowski allowed for a straightforward generalization and that was the path followed by 
Einstein (1879 - 1955) from 1907 onward after realizing that Newtonian gravity did not fit within the framework of special 
relativity. Later on, in collaboration with his friend and Zurich's Technical University colleague, Marcel Grossmann (1878 - 1936), 
Einstein wrote a seminal paper, albeit not quite consistent, in 1913, where it was clearly spelled out that 
Riemannian geometry was actually the most general and natural setting for physics (see, for instance, Ref. \cite{Pais} for a detailed 
account). Of course, these developments relied strongly on the XIX century work of Lobatchevski(1793 - 1856), Bolyai (1802 - 1866), 
Riemann (1826 - 1846) and Gauss (1777 - 1855), who have shown that flat spaces  
are a particular case of a much wider class of spaces with non-vanishing 
curvature. This was indeed a quite new idea, even though, space and time of day to day affairs, 
were still regarded as  
{\it a priori} concepts that preceded all experience and were independent of 
any physical phenomena in the Newtonian (and Kantian) sense. But, of course, general relativity revolutionized this view 
showing that space and time are actually associated with a given energy-matter distribution, so that 
the Newtonian perspective was, at best, just an approximation 
to the inner nature of space and time.  

Actually, the notion of world-lines in general relativity is basically similar to the one 
in special relativity, with the difference that in the former, space-time can be curved. 
The dynamics of the metric is determined by the Einstein field 
equations and depends on the energy-mass distribution in space-time. As before, 
the metric defines light-like (null), space-like and time-like curves. Also, 
in general relativity, world lines are time-like curves in space-time, where 
time-like curves fall within the light-cone. However, a light-cone is not 
necessarily inclined at 45 degrees to the time axis, if one adopts the unit system where $c=1$. 
However, this is an artifact of the chosen coordinate system, and reflects the coordinate 
freedom, the very essential diffeomorphism invariance of general relativity. Any time-like 
curve admits a co-moving observer whose ``time axis" corresponds to that 
curve, and, since no observer is privileged, we can always find a local 
coordinate system in which light-cones are inclined at 45 degrees to the 
time axis. Furthermore, the world-lines of free-falling particles or objects, such as the ones of planets around 
the Sun or of an astronaut in space, are minimal length curves, the so-called geodesics.

However, one should keep in mind that general relativity contains geometries which defy the very core of physical reasoning. 
Indeed, the existence of {\it closed time-like curves} contradicts the essential features of causality and chronology. Moreover, 
singularities, unavoidable in general relativity, once closed time-like curves are absent and geometry is set by well-behaved 
matter-energy configurations, imply that geodesics cannot exist in the whole space-time. 

The referred condition on matter-energy is fairly specific as in the Hawking-Penrose singularity theorems and is 
tied up with the physical nature of a manifold \cite{HEllis}. A Lorentzian manifold $(M, g)$ is said to be 
physically well-behaved if it satisfies the {\it strong energy condition}:
\begin{equation}
\label{Strongenergy} R_{\mu \nu} V^{\mu} V^{\nu} \ge 0~,
\end{equation}  
for any time-like vector field, $V^{\mu}$. From Einstein's equations this statement is equivalent, 
for more than two $d$-spatial dimensions, to the condition on the energy-momentum tensor and its trace, $T$, 
\begin{equation}
\label{EMtensor-strong} T_{\mu \nu} V^{\mu} V^{\nu} \ge {T \over d-1}  V_{\mu} V^{\mu}~. 
\end{equation}  
This condition is fulfilled by spaces dominated by the vacuum with a positive cosmological constant ($\Lambda \ge 0$), 
and by a perfect fluid if $\rho + 3 p \ge 0$, where $\rho$ and $p$ correspond to the energy density and isotropic pressure, 
respectively. 

Of course, the fundamental assumption here is the connection between physical spaces with mathematical spaces that satisfy the 
Einstein field equations. A generic mathematical space is fundamentally described by a 
$d$-dimensional differentiable manifold $M$ endowed with a symmetric, non-degenerate second-rank tensor, the metric, $g$. 
A manifold under these specifications is said to a pseudo-Riemannian manifold, $(M, g)$, as it has a Lorentzian signature  
$(-, +, ..., +)$ - it is Riemannian if it has signature $(+, ..., +)$. A  differentiable 
manifold admits a Lorentzian signature if it is noncompact or has a vanishing Euler characteristic.
The Italian mathematician Tullio Levi-Civita (1873 - 1941), showed a well known theorem according to which 
a pseudo-Riemannian manifold has a unique symmetric affine connection compatible with the metric, 
being hence equipped with geodesics. 

The spaces relevant to physics correspond to solutions of the 
Einstein equations with a cosmological 
constant\footnote{The ``natural'' system of units is used: $c=\hbar = k_{B}=1$.}, $\Lambda$:
\beq
R_{\mu \nu} - {1 \over 2}  g_{\mu \nu} R + \Lambda g_{\mu \nu}  = 8 \pi G T_{\mu \nu}~,
\eeq
where $R_{\mu \nu}$ is the Ricci curvature of $M$, $R$ its trace, $G$ is Newton's constant 
and $T_{\mu \nu}$ is the energy-momentum tensor of matter in $(M, g)$.

Thus, through Minkowski unification, it was possible to intimately relate physics back to geometry, a connection quite 
dear to Galileo (1564 - 1642) and Newton, but somewhat lost in the XIX century physics. In a 
previous text by this author \cite{Bertolami06b}, this methodology was referred to as ``C\'ezanne's principle'', given 
the suggestive connection it has with the writings of the French painter 
who first reflected on the then new cubist revolution. According to C\'ezanne (1839 -1906), the essence of the new movement 
was the description of nature through purely geometrical forms.
One could further mention that the relativity revolution is seen by many thinkers to be somewhat similar to the one 
that took place in modern art through 
movements like futurism, cubism, and other ``isms'' which have shown to be possible to depict in a single plane various points of view, 
as well as the mutation of reality through superimposing images. In this way, time was introduced into the arts that were 
traditionally associated with space, such as sculpture, architecture and painting, in opposition to the ones associated with 
time, music and literature.

The fundamental insight of Minkowski allowed for the generalization of the special theory of relativity and its 
application for understanding the inner secrets of the microscopical world through the development of quantum field 
theory. As far as experimental evidence allows us to unravel, quantum field theory methods are consistent down to about $10^{-18}$ m. 
On the largest known scales, general relativity, where the most general space-time continuum is actually 
curved, cosmological evidence seems to fit the so-called cosmological standard model up to the horizon size, 
i.e. distances up to $10^{26}$ m (see e.g. \cite{Will,BPT}). This is an impressive vindication of Minkowski's World Postulate.

\section{The mystery of Kronos}
\label{sec:2}

If at the physical and conceptual level, one could assume that most of the merit of the Minkowski unification 
is due to the radically different 
view implied by special relativity a few years earlier thanks to the work of Einstein, one should realize that 
from the historical and philosophical  
standpoint, the proposal of Minkowski is a remarkable culmination of more than two thousand years of research about the nature of space 
and time. In what follows we shall discuss some of the most conspicuous philosophical ideas about the space and time 
problematic. Our main sources for the ensuing discussion are Refs. 
\cite{Russell,Abbagnano,Coveney,Montheron,Ellis,Bertolami92,Bertolami06b,Lobo}.   

The very first manifestations of articulate rational thinking about the origin of the world, 
myths of creation, often regarded space and time as inseparable, as the original 
creation act gave birth to both space and time - actually, likewise the modern theory of the Big-Bang. In ancient Hellenic period, 
space and time were seen as two essential features of 
reality, but in many instances, regarded as distinct entities. 

Indeed, at first, space and time seem to be quite different. Space can be freely experienced as one can move in any
direction without restriction. Time however, has a well defined direction. Past and future are clearly distinct as our action 
can affect only the latter. We have memory, but not precognition. Matter, organic or otherwise, tends to decay rather than to organize itself 
spontaneously. There seems to exist at least 3 distinct spatial dimensions\footnote{The Finnish physicist 
Gunnar Nordstr\"om (1881 - 1923) was the first to speculate in 1909, that space-time could very well have more 
than four dimensions. A concrete realization of this idea was put forward by Theodor Kaluza (1885 - 1954) 
in 1919 and Oskar Klein (1894 - 1977) in 1925, 
who showed that an unified theory of gravity and electromagnetism could be achieved through a 5-dimensional (4-spatial 
and a single time dimension) version of General Relativity. These extra dimensions in order to have passed undetected 
can be either compact and very small or very large if the known fundamental interactions, excluding gravity, can test only 3-spatial 
dimensions. In any case, the extension of the number of spatial dimensions has been widely considered in attempts to unify all known 
four interactions of nature. For instance, the requirement that supersymmetry is preserved in 4 dimensions, 
from the original $10$-dimensional superstring theory, implies that 6 dimensions of the world are compact \cite{CHSW}. 
Connecting all string theories through $S$ and $T$ dualities suggests the existence of an 
encompassing theory, M-theory, and that space-time is $11$-dimensional \cite{Witten95}.}  
while there is only a single time dimension. Actually, the fact that in many physical theories 
the time dimension is just a parameter turns it into an ``invisible dimension'' \cite{Prigogine1}. Notice that if the number 
of time dimensions is greater than one, one expects all type of complications as, on quite general grounds, the Partial 
Differential Equations that describe the physical phenomena are ultra-hyperbolic, which leads to unpredictability, or   
in weird ``backward causality'' (see e.g. Ref. \cite{Bertolami06a} for a discussion).  

Let us resume the philosophical discussion. 
Space has always been regarded as the arena of all manifestations of nature. 
Everything lies in space and the intrinsic and fundamental 
relationships between the most basic elements of {\it everything} could be decomposed into points, straight lines and
geometrical figures in two or three dimensions and whose properties were monumentally described by Euclid's (ca. 330 - 275 B.C.) 
geometry. These relationships would in turn reveal the intrinsic properties of space itself. 
Of course, the fundamental role of space in the Hellenic philosophical thinking was more than evident on the speculative thinking 
of the pre-Socratic Zeno (495 - 435 B.C.) and Pythagoras (ca. 569 - 500 B.C.) 
and many others after them. For instance, for Plato (428 - 349 B.C.), ``God ever geometrizes''. 
For Aristotle (384 - 322 B.C.), the 
``geometrical'' method and proof was the intellectual reasoning model that should be used in natural sciences, ethics, 
metaphysics and so on.    

There were however, instances where philosophical thinking hinted at a hidden connection between space and time. 
For instance, for Zeno, the Dichotomy, the Achilles - tortoise, and the Arrow 
paradoxes stressed the fundamental difficulty in reconciling motion, that is dislocation in space, actually along a straight line, 
with the concepts of continuity and divisibility. In his Stadium paradox, Zeno considers three rows of bodies lying on a 
line and how the opposite relative motion of two of the rows ``proof that half the time may be equal to double the time'' \cite{Bell}. 
Of course, these puzzles reflect the immaturity of the mathematical thinking at Zeno's time 
concerning the infinitesimal. But, it is   
suggestive that Zeno was already seeing that paradoxes in space and time were related in the real physical 
world through motion. A contemporary physicist could not fail to see that the divisibility process considered by Zeno 
could not go on indefinitely, as the 
fundamental limitation of quantum effects on the fabric of space-time would arise at Planck length, $10^{-35}$ meters (or 
equivalently Planck time level, $10^{-44}$ seconds), the length where the Schwarzschild radius of a particle equals 
its Compton wavelength\footnote{Actually, in some quantum gravity approaches, as for instance in loop quantum gravity, 
space-time is suggested to have, at 
its minutest scale, presumably the Planck scale, $L_P \simeq 10^{-35}$ m, 
a discrete structure \cite{ARSmolin}. In superstring/M-theory, the space-time continuum is an emergent property that arises from 
the ground state excitations of closed strings, one of the fundamental objects of the theory.}.

For Pythagoras, who was the first to understand that above the application of mathematical 
tomb rules stood the {\it proof} of the fundamentals behind the rules, one could argue that 
the association of mathematics with music implied an   
inner connection between geometry and {\it tempo}, the very essence of music, time. 

Actually in the Hellenic mythology, more specifically in the Orphic cosmogonies, time had a particularly interesting standing. 
Khronos, the primeval god of time, emerged as a self-formed divinity at the beginning of creation \cite{Orphica}:

\noindent
``Originally there was Hydros (Water) and Mud, from which Ge (Earth) formed solidified ... The third principle after the Hydros and Ge 
was engendered by these, and was a Drakon (Serpent) with extra heads of a bull and a lion and a god's countenance in the middle; it 
had wings upon its shoulders, and its name was Khronos (Unaging Time) and also Herakles. United with it was Ananke (Inevitability, 
Compulsion), being of the same nature, her arms extended throughout the universe and touching its extremities ...''  

Thus, Khronos and Ananke encircled the cosmos from the time of creation, 
and their passage drives the circling of heaven and the eternal flow of time. 
 
>From myth to rational thinking, time was insightfully dissected by Heraclitus (ca. 535 - 475 B.C.) 
and by Aristotle (384 - 322 B.C.). 
Heraclitus understood the world as a unit resulting from diversity in eternal transformation. 
Time is what allows events to occur as a result  
of a web of antinomies. The formulas:''You cannot step twice in the same river; for fresh waters are ever flowing in upon you'' 
and ``The sun is new every day'', capture the powerful idea that ``all things are flowing''.  

For Aristotle, ``time is a measure of motion according to the preceding and the succeeding'', time is associated to the evolution, 
change in quantity and/or quality, of all occurrences in nature. For the most influential disciple of Plato, 
time is intimately related to motion, and with the counting process that is, with numbers. Aristotle's view leads 
to an operational connection between time and any material system that 
can be used as a standard for measuring the passage of time: clepsydras, sand clocks, sun clocks, pendulums and clocks. Since 
ancient times, the motion of earth around the sun has been the measure of day to day activities. Thus, in essence, 
our closest connection to time is the very one put forward by Aristotle more than two thousand years ago. 
This is irrespective of any technological 
development, whether we use the regularity of the astronomical motions in the solar system (which, of course, 
cannot be exact due to the complexity of the physics 
behind these motions) or atomic clocks, or even binary pulsars, possibly the most precise clocks in the universe. 
In practical terms, one defines the second, the fundamental unit of time, as 
$1/86400$ of the duration of the average solar day, or $9,192,631,770$ periods of transition the radiation corresponding 
to the transition between the two hyperfine levels of the ground state of the cesium-133 atom\footnote{This definition concerns 
a cesium atom at rest at a temperature of $0 K$, such that the ground state is defined at zero magnetic field.}.

The average solar motion is defined in terms the of the idealized uniform motion of the sun 
along the celestial equator. The difference 
between this idealized motion and the real motion is called ``equation of time''. 

Coming back to the philosophic discussion, the vision of time by Saint Augustine (354 - 430) is remarkable in its modernity. 
He believed that 
the origin of time was the creation of the world. This was a fundamental precaution since God's eternity made 
its identification with time impossible. In his {\it Confessions} he expresses his view that the notion of 
present is the most fundamental feature of time:

\noindent
``The present of things past is memory. The present of things at present is perception. 
The present of things in the future is expectation.''

Another fundamental aspect of Saint Augustine's view of time is his rejection of the doctrine of cyclic history. In many ancient 
civilizations, Sumer, Babylon, Indian, Mayan, 
the regular patterns of tides, seasons and the cyclic motion of heavenly bodies entailed from the fact that time 
itself was circular. Day is followed by night, night by day, summer follows winter, winter by summer, old moon follows new, new one by old one, and so why not history itself ?

The cyclic temporal pattern was a noticeable feature of Greek cosmology. The Stoics believed that on every 
instance after the planets returned 
to their exact relative positions as at the beginning of time, the whole cosmos would become renewed. Nemesius, 
Bishop of Emesa says at the Vth century: ``Socrates and Plato and each individual man will live again ... And this 
restoration of the universe takes place not once, but again and again, indeed to all eternity without end''. 
The Maya civilization of Central America believed that history would repeat itself every 260 years, a period of fundamental importance 
in their calendar. The ancient Indians (Hindus, Budhists, Jains) extended the idea of a Great Year, a full cycle, into a hierarchy of 
Great Years. The destruction and re-creation of individuals and creatures occurred in a day of Brahma. A day of Brahma lasted 4 thousand 
million years. The elements themselves and all forms will undergo a dissolution into Pure Spirit, which then incarnates itself 
back into matter every lifetime of Brahma, that is, about $311 \times 10^{12}$ years. The lifetime of Brahma is the longest cycle and is 
repeated {\it ad infinitum} (see e.g. in \cite{Tipler} and references therein).  

The notion of cyclic time was thus regarded more comfortable, 
as a time arrow would mean instability, and inevitably irreversible 
change. The {\it myth of eternal return} was a central idea of many ancient civilizations, exceptions being the Hebrews and 
the Zoroastrian Persians. Thus, it was through the Judaeo-Christian tradition and most definitely through Saint Augustine 
that linear and irreversible time got established in the Western culture. According to 
Saint Augustine, Christ's death and Crucification was an unique event, and from then on the cultural prominence of the Roman Christian 
Church took charge of ``spreading the word'' about the linear nature of time and history. But even before that it was 
considered heresy to claim otherwise and transgressors were being punished in an exemplary way. Despite that, defenders 
of the doctrine of circular time, sometimes referred to as the {\it annulars}, 
were not short of conviction. In the III century, Euforbo, a presumed member of the {\it annulars} sect, while burnt at the stake, 
is alleged to have 
screamed: ``This happened already and will happen again. You do not light a pyre, you light a labyrinth of fire. If all 
the pyres that I have been were put together they would not fit on earth and would blind the angels. I have said that many times'' 
\cite{Borges}.      

In this context, it is particularly interesting to remark that Spinoza (1632 - 1677), 
who believed that everything could be attributed to 
a manifestation of God's inscrutable nature, and as such occurred by absolute 
logical necessity, defended the idea that time was unreal, and hence all emotions associated to an event in the 
future or in the past are 
contrary to reason. For this philosopher, whose ethics should follow as in Euclid's geometry from definitions, axioms and theorems, 
it should be understood that ``in so far as the mind conceives a thing under the dictate of reason, it is affected equally, 
whether the idea be of a thing present, past or future''. 

The existence of physical time was also doubted by the 11th century, Persian philosopher Avicenna, who argued that time 
exists only in the mind due to memory and expectation. It is remarkable that similar views were expressed by Einstein himself. 
Indeed, in a letter to his life long friend, Michelle Besso, he writes (see e.g. \cite{Prigogine1,Prigogine2}:

\noindent
``There is no irreversibility in the basic laws of physics. You have to accept the idea that subjective time with its 
emphasis on the now has no objective meaning''.

Later, on the occasion of Besso's passing away, in a letter addressed to his widow and son, he says:

\noindent
``Michelle has preceded me a little in leaving this strange world. This is not important. For us who are convinced physicists, 
the distinction between past, present and future is only an illusion, however persistent''. 

Also worth mentioning is the view of time of one of the most brilliant opponents of Newton, Gottfried Leibniz (1646 - 1716). 
He argued that time cannot be an entity existing independently of actual events. For Leibniz absolute space does not exist. Space 
is the relative configuration of bodies that exist simultaneously. Thus, time is the succession of instantaneous configurations, 
and not a flux independent of the bodies and their motion. It follows that as time concerns a 
chronology of events, in a universe where 
nothing happens, there is no time. This disagreement with the very basis of Newtonian mechanics lead Leibniz to suggest that 
mechanics should be built strictly in terms of observed elements. This view was shared by the science philosopher 
Ernest Mach (1838 - 1916) and 
Heinrich Rudolph Hertz (1857 - 1894), who actually developed a ``relational'' mechanics based on the ideas of Leibniz. It is well 
known that Einstein was particularly impressed by Leibniz's ideas on space and time and these inspired him when constructing the 
general theory of relativity, and also played an important role in his life long rejection of quantum mechanics.     

Causation, and thus time ordering, is according to Hume (1711 - 1776) the basis of human 
understanding. He warns that: ``We ought not to receive as reasoning any of the observations we make concerning identity, 
and the relations of time and place; since in none of them the mind can go beyond what is immediately present to the senses ...
Causation is different in that it takes us beyond impressions of our senses, and informs us of unperceived existences.'' 
It is arguable whether on purely philosophical terms Hume's doctrine stands on its own; however the decomposition of human 
perception down to the physiology of nervous tissues, down to its chemistry and then primarily the causal character of the 
physical laws, render Hume's proposition quite plausible.

In the {\it Critic of Pure Reason}, first published in 1781, Immanuel Kant (1724 - 1804) advances ideas 
about space, time and actually the main metaphysical problems of his time that turned out to be particularly influential. 
For Hume, the law of causality is not ``analytic'', that is, a proposition in which 
the predicate is part of the subject, such as for instance, an ``equilateral triangle is a triangle''. Kant agreed in that 
causation was a crucial starting point, however for him this law is synthetic and known {\it a priori}. A ``synthetic'' 
proposition is the one that is not analytic. All propositions that we know only through experience are synthetic. An ``empirical'' 
proposition is one which we cannot know except through the sense-perception, either our own or that of someone's testimony. 
So are the facts of history and geography, as well as the laws of 
science in so far as our knowledge of their truth depends on observational data. 
An a ``priori'' proposition, on the other hand, although susceptible of elucidation by experience, has, after inspection, a basis 
other than experience. All the propositions of pure mathematics are {\it a priori}. Kant then poses the question: How are 
synthetic judgments a priori possible ?  His solution can be expressed in the following way: The outer world can only 
excite our senses, but it is our own mental apparatus that orders our sensations in space and time, providing in this way the 
means through which we understand experience. Space and time are thus subjective, they are part of our apparatus of perception; 
however, precisely because of their a priori nature, whatever we experience will exhibit the features that can be dealt with 
through geometry and the science of time. Space and time, Kant argues, are not concepts; they are means of intuition, forms of viewing 
or looking at the world.   

Of some interest to physics\footnote{One should keep in mind that in 1755 Kant anticipated, in his {\it General 
Natural History and Theory of the Heavens}, how from Newton's mechanics one could explain the origin of the 
solar system. This nebular hypothesis was actually made mathematically plausible by Laplace's (1749 - 1827) many decades later.} 
is also the part of the {\it Critic of Pure Reason} which deals with the fallacies, the ``antinomies'', 
mutually contradictory propositions which can be both proved to be true. They 
that arise from applying space and time or the categories (things in themselves) to what cannot be 
experienced. Kant discusses four of such antinomies, each consisting of thesis and antithesis. The first states: ``The world 
has a beginning in time, and is also limited as regards to space''. The antithesis says: ``The world has no beginning, 
and no limits in space; it is infinite in regard to both time and space.'' 
The second antinomy proofs that every composite substance is both, and is not, made up of simple parts. The third antinomy states that 
there are two kinds of causality, one associated to the laws of nature, the other concerning that of freedom. 
The antithesis maintains that there is only the causality related to the laws of nature. Finally, the 
fourth antinomy shows that there is, and there is not, an absolute necessary Being. In a subsequent section, Kant destroys all 
purely intellectual proofs of the existence of God, even though he clarifies that he has other reasons for believing in God. 

The antinomies have greatly influenced another important German philosopher, George Friedrich Hegel (1770 - 1831).      
Hegel believed in the unreality of the separateness, whether atoms or souls. The world is not a collection of self-subsistence 
units; nothing, Hegel held, is ultimately and completely real except the whole. 
Related to this is his disbelief in the reality of space 
and time as such, as these, if taken as completely real, involve separateness and multiplicity, which he regards as an illusion or 
as a mystic insight. Wholeness is the reality and this is rational as the rational is real. The engine of his metaphysical 
view of the world was dialectic: a thesis, antithesis and synthesis which sets consistency with the whole. His dialectic 
method applied to history in his {\it Philosophy of History}, could arguably give unit, and meaning 
to revolutions and movements of human 
affairs at the level of ideological currents of thought. Atributable to Karl Marx (1818 - 1883) is 
the theory that actually, the ultimate cause of human affairs 
moves dialectically, due to the clash of conflicting means of economical production. It is interesting 
that Marx regarded his insight about the development of human society as being analogous to the teleological evolution 
of species, the engine and clock of biological change, as first proposed by Darwin (1809 - 1882) and Alfred Russel Wallace 
(1823 - 1913) in 1858.     

For Henri Bergson (1859 - 1941), intelligence and intellect can only form a clear idea of discontinuity and immobility, 
being therefore unable to understand life and to think about evolution. Intellect tends to 
represent {\it becoming} as a series 
of states. Geometry and logic, the typical products of intelligence, are strictly applicable to solid bodies, but to everything 
else reasoning must be checked by common sense. Actually, Bergson believes 
that the genesis of intelligence and the origin of material bodies are 
correlative and have been developed by reciprocal adaptation. For him, the growth of matter and intellect are simultaneous. 
Intellect is the power of seeing things as separate and 
matter is what is separated into distinct things. However, in reality, there are no 
separate solid things, only a continuous stream of becoming. Becoming being an ascendant movement that leads to life, or a descendant 
movement leading to matter.

For Bergson, space and time are profoundly dissimilar. The intellect is associated with space, while instinct and intuition are 
connected with time. Space, the characteristic of matter, arises from a dissection of the flux which is really illusory, although 
useful in practice. Time, on the contrary, is the essential feature of life or mind. ``Wherever anything lives, there is, open 
somewhere a register in which time is being inscribed''. But time here is not a ``mathematical'' time, an homogeneous assemblage 
of mutually external instants. Mathematical time, according to Bergson, is actually a form of space; the time which is the essence 
of life is what he refers to as {\it duration}. ``Pure duration is the form which our conscious states assume when our ego lets 
itself live, when it refrains from separating its present state from its former states''. Duration unites past and present into an 
organic whole, where there is mutual entanglement and succession with distinction.

We could not draw to an end this brief account on the philosophical thinking about the nature of space and time without a reference  
to the images that often arise in poetry and literature, where time in particular, is quite often insightfully evoked. 
Indeed, from the Rub\'aiy\'at of Omar Khayy\'am (1048 - 1131) to ``La recherche du temps perdu'' of Marcel Proust (1871 - 1922), 
from William Blake (1757 - 1827) to contemporary authors such as Imre Kert\'esz and Paul Auster, time and memory are central themes 
in the literary context. One often finds quite profound glimpses on the nature of time. In ``Du c\^ote de chez Swann'', Proust 
says: 

\noindent
``The past is hidden beyond the reach of intellect, in some material object (in the sensation that the object will give us). 
And as for that object, it depends on chance whether we come upon it before we ourselves die.''     
   
Actually, this fundamental impression which allows us to expand our imagination and build theories based on 
the discovery of these ``time capsules'', a rock containing a fossil, the light of a distant star, the cosmic microwave background 
radiation or an ancient picture. 

Time, the cycles of life and the hope of prevalence as put forth by Shakespeare (1564 - 1616) in his LX sonnet \cite{Shakespeare}:

\vspace{0.1cm}

\noindent
Like as the waves make towards the pebbled shore,

\noindent
So do our minutes hasten to their end;

\noindent
Each changing place with that which goes before,

\noindent
In sequent toil all forwards do contend.

\noindent
Nativity, once in the main of light,

\noindent
Crawls to maturity, wherewith being crown'd,

\noindent
Crooked eclipses 'gainst his glory flight,

\noindent
And Time, that gave, doth now his gift confound.

\noindent
Time doth transfix the flourish set on youth,

\noindent
And delves the parallels in beauty's brow;

\noindent
Feeds on the rarities of nature's truth,

\noindent
And nothing stands but for his scythe to mow:

And yet to times in hope my verse shall stand,

Praising thy worth, despite his cruel hand.

\section{Arrows of time}
\label{sec:3}

By the second half of XIX century, the development of the kinetic theory of matter by Maxwell (1831 - 1879), 
Clausius (1822 - 1888) and Boltzmann (1844 - 1906) revived once again the discussion on 
the dichotomy between the linear evolution of time and the eternal recurrence of motion.   

The idea of a cyclic time and of an eternal return was recovered in philosophy by Herbert Spencer (1820 - 1903) and 
Friedrich Nietzsche (1844 - 1900) about the same time that Poincar\'e (1854 - 1912) showed his well known recurrence theorem. 
For sure, their ``proofs'' cannot be considered rigorous by the standards of physics and 
mathematics; however, interestingly, the ``proof'' of Nietzsche 
contains elements which can be regarded as relevant for any discussion of the issue, 
such as a finite number of states, finite energy, no creation of the universe and 
chance-like evolution. In his {\it Dialectic of Nature}, the philosopher and revolutionary politician, companion and 
co-author with Karl Marx of the {\it Communist Manifesto}, Friedrich Engels (1820 - 1895) wrote in 1879:

\noindent
`` ... an eternal and successive repetition of worlds in an infinite time is the only logical conclusion of the coexistence of 
countless worlds in an infinite space ... It is in an eternal cycle that matter moves itself. ``  

Let us now turn to the physical discussion. Newton's equations have no intrinsic time direction, 
being invariant under time reversal; however, Poincar\'e showed in 1890, in the context of 
classical mechanics, a quite general recurrence theorem, according to which any isolated system, which includes 
the universe itself, would return to its initial state given a sufficiently long time interval.  

Poincar\'e's theorem is proved to be valid in any space $X$ on which there exists a one parameter 
map $T_i$ from sets $[U]$ and a 
measure $\mu$ on $X$ such that: i) $\mu(X)=1$ and ii) $\mu(T_{t_0}(U))=\mu(T_{t_0+t}(U))$ for any subset of $X$ and any $t_0$ and $t$. 
In classical mechanics, condition i) is ensured by demanding that space $X$ is the phase space of a finite energy system in a 
finite box. If $\mu$ is the distribution or density function, $\rho$, in phase space and $T_t$ 
is the evolution operator of the mechanical 
system (associated with the Hamiltonian or Liouville operator), then condition ii) follows from Liouville's theorem: $d\rho /dt=0$. 
It thus follows that classical mechanics is inconsistent with the Second Principle of Thermodynamics.  

Of course, the recurrence issue was a key concern to Boltzmann, who in the 1870s realized that deducing 
irreversibility, an arrow of time, from the mechanics of 
atoms was impossible without using averaging arguments. It was in the context of his efforts to understand the statistical 
equilibrium with the Liouville equation 
that he obtained in 1872 a time-asymmetric evolution equation, now referred to as Boltzmann equation, 
satisfied by a single-particle distribution function of a molecule in a diluted gas. From this he could construct 
a mathematical function, the so-called ${\cal H}$-function which is a strictly decreasing function of time. Identifying the 
${\cal H}$-function with entropy with minus sign, Boltzmann could claim to have solved the irreversibility problem at molecular 
level.

However, in order to arrive at his result Boltzmann had to rely on the ``molecular chaos hypothesis'' (Stosszahlansatz), i.e. 
on the assumption that molecules about to collide are uncorrelated, but following the collision they are correlated as their 
trajectories are altered by the collision. Ernest Zermelo (1871 - 1953), young assistant of Planck (1858 - 1947) 
in Berlin, and Johann Josef Loschmidt (1821 - 1895), 
friend of Boltzmann, argued that the time-asymmetry obtained by Boltzmann was entirely due to the 
time-asymmetry of the molecular chaos assumption. Twenty years later, Zermelo attacked Boltzmann once again, now armed with 
Poincar\'e's recurrence theorem. Boltzmann attempted to save his case through a cosmological model. He suggested 
that as a whole the universe had no time direction, but rather individual regions could be time-asymmetric when through
a large fluctuation from equilibrium it would yield a region of reduced entropy. These low entropy regions would evolve back to 
the most likely state of maximum entropy, and the process would repeat itself in agreement with Poincar\'e's theorem.      

Having become clear that a finite system of particles would be recurrent and not irreversible in the long run, Planck considered 
whether irreversibility could emerge from a field theory such as electromagnetism. The point was to derive irreversibility 
from the interaction of a continuous field with a discrete set of particles. Starting to tackle the problem 
in 1897 in a series of papers, 
the work of Planck culminated with his discovery of the quantum theory of radiation in 1900. From Planck's arguments, 
Boltzmann remarked 
that as a field can be regarded as a system with an infinite number of degrees of freedom, and hence expected to be analogous 
to a mechanical system with an infinite number of molecules, an infinite Poincar\'e recurrence period and thus agreement with the observed irreversibility and the Second Principle would follow. 

However, the persistent objections of influential opponents such as Ernest Mach and Friedrich 
Ostwald (1853 - 1932), led Boltzmann into depression and a first suicide 
attempt while at Leipzig before assuming his chair in Vienna in 1902. The intellectual isolation, as he was the sole survivor 
of the triumvirate of theoreticians along with Clausius and Maxwell, who had developed the kinetic theory of matter, and the 
continuous deterioration of his health led him to suicide and death at Duino, a seaside holiday resort on 
the Adriatico coast near Trieste, on the 5th September 1906. He was 62 years old. Boltzmann death is even more tragic 
when one realizes that it happened on the very eve of the vindication of his ideas.      
    
However, the irreversibility problem has somehow resisted a straightforward answer. In 1907, the couple Ehrenfest, 
Paul Ehrenfest (1880 - 1933) and Tatyana Afanasyeva (1876 - 1964), 
(see e.g. Ref. \cite{Huang}), further developed Boltzmann's idea of averaging over a certain region $\Delta$, 
of the phase space and showed 
that the averaged ${\cal H}$-function would remain strictly decreasing in the thermodynamical 
limit, after which $\Delta$ could be taken as 
small as compatible with the uncertainty principle.

In 1928, Pauli (1869 - 1958) 
considered the problem of transitions in the context of quantum mechanical perturbation theory and showed that 
consistency with the Second Principle of Thermodynamics would require a ``master equation'':

\beq
{d p_i \over d t} = \sum_j ~(\omega_{ij} p_j -\omega_{ji} p_i) ~,
\eeq  
where $\omega_{ij}$ is the conditional probability per unit of time of the transition $j \goto i$ and $p_i$ is the probability of 
state $i$. Assuming the {\cal H}-function to be given by 
 
\beq
{\cal H} = \sum_i ~p_i \ln p_i ~,
\eeq   
it follows that ${d {\cal H} \over d t} \leq 0$. This approach is quite suggestive as it indicates, as stressed by Boltzmann, 
that irreversible phenomena should be understood in the context of the theory that best describes microscopic physics. 

More recently, Prigogine (1917 - 2003) and collaborators put forward a more radical approach, 
according to which irreversible behaviour 
should be already incorporated in the microscopic description (see Ref. \cite{Prigogine1} for a pedagogical discussion). In 
mathematical terms the problem amounts to turning time into an operator which does not commute with the Liouville operator, 
the commutator of the Hamiltonian with the density matrix. In physical terms, this proposal implies that the 
reversible trajectories cannot be used, leading to an entropy-like quantity which is a strictly increasing function of time.

But, if the problem of explaining the irreversible behaviour of all macroscopic systems from microphysics is already 
quite difficult, one should realize that there exists in nature quite a variety of phenomena whose behaviour indicate an immutable 
flow from past to present, from present to future. The term ``arrow of time'', already used in text, was coined by 
the British astrophysicist and cosmologist Arthur Eddington (1882 - 1944) \cite{Eddington} to characterize this evolutionary behaviour.  
Let us briefly describe these phenomena:

\vspace{0.4cm}

\noindent
1) The already discussed time asymmetry inferred from the growth of entropy in irreversible and dissipative phenomena, as described 
by the Second Law of Thermodynamics.

\vspace{0.4cm} 

\noindent
2) Nonexistence of advanced electromagnetic radiation, coming from the infinite and converging to a source, even though solutions 
of this nature are legitimate solutions of the Maxwell's field equations.

\vspace{0.4cm} 

\noindent
3) The collapse of wave function of a quantum system during the measurement process and the irreversible 
emergence of the classical behaviour, even though the fundamental equations of quantum mechanics and statistical 
quantum mechanics, Schr\"odinger's and Von Neumann's equations, 
respectively, are invariant under time inversions for systems described by 
time-independent Hamiltonians (see e.g. Ref. \cite{Penrose} for a vivid discussion).

\vspace{0.4cm}

\noindent
4) The exponential degradation in time of systems and the exponential growth of self-organized systems (given a sufficiently large 
supply of resources). In the development of self-organized systems, a particularly relevant role is played by complexity. 
The fascinating aspects of phenomena in this context has lead authors to refer to them as ``creative evolution'', ``arrow of life'', 
``physics of becoming'' \cite{Coveney,Montheron,Prigogine1,Prigogine2}. In these discussions, the chaotic 
behaviour plays an important role 
given that complex systems are described by non-linear differential equations. This chaotic behaviour gives 
origin to an extremely rich spectrum of possibilities for describing self-organized systems as well as a paradoxically predictable 
randomness as chaotic branches are deterministic (see for instance, Refs. \cite{Coveney,Gleik}).

\vspace{0.4cm} 

\noindent
5) The discovery of the CP-symmetry violation in the $K^0 - \bar{K}^0$ system implies, on account of the CPT-theorem, that 
the T-symmetry is also violated. This means that on a quite elementary level there exists an intrinsic irreversibility. The violation 
of the CP-symmetry and also of baryon number in an expanding universe are conditions from which the observed baryon asymmetry 
of the universe can be established (see e.g. \cite{Buchmuller} and references therein). An alternative way to achieve 
the baryon asymmetry of universe is through the violation of the CPT-symmetry (see {\cite{Bertolami97} and references therein).

\vspace{0.4cm}

\noindent
6) Psychological time is clearly irreversible and historical. The past is recognizable, while the future is open. 

\vspace{0.3cm}

\noindent
7) The so-called gravito-thermal catastrophic behaviour \cite{Lynden-Bell} of systems bound gravitationally, 
implies, given their negative specific heat, that their entropy grows as they contract beyond limit. 

\noindent 
On the largest known scales, the expansion of the Universe, which is adiabatic, is a quite unique event, and as such, 
is conjectured to be the master arrow of time to which all others are subordinated.

\vspace{0.3cm}

\section{Open questions}
\label{sec:4}

Let us briefly discuss in this section a few problems concerning the nature of space-time that remain unsolved. These include 
the issue of a putative correlation between the above described arrows of time, the problem of nonexistence of an explicit time in 
the canonical Hamiltonian formulation of quantum gravity, the question of solutions of general relativity and other gravity theories 
that exhibit closed time-like curves and whether the universe evolves after all in a cyclic way.

\subsection{Are the arrows of time correlated ?}

The existence of systems, from which a time direction can be inferred, is not on its own very surprising, as it is in the 
core of all dissipative phenomena. One could argue that this property reflects, for instance, a particular choice of boundary 
conditions which constrain the state of the universe, rather than any restriction on its dynamics and evolution.
However, this point of view cannot account for the rather striking fact that the known arrows of time seem all to point in the
same time direction. In the following, we shall briefly overview some of the ideas put forward to relate the arrows of time.
Extensive discussions can be found in Refs. \cite{Davies,Zeh,GellMann}.

In his book {\it The Direction of Time}, the philosopher Hans Reichenbach (1891 - 1953) 
\cite{Reichenbach} argued in a rather circular way that 
the arrow of time in all macroscopic phenomena has its origin in causality, which in turn should be the origin of 
the growth of entropy. In 1958, the cosmologist Thomas Gold (1920 - 2004) put forward the remarkable idea that all arrows of time 
should be subordinated to the expansion of the universe \cite{Gold}. This speculation gave origin to demonstrations, although 
not quite entirely successful, that the propagation of the electromagnetic radiation was indeed related to the expansion 
of the universe \cite{Hogarth,Hoyle}. The problem is that the obtained solutions are somewhat puzzling. Indeed, it is found 
that: retarded radiation is found to be compatible only with a steady-state universe, while advanced radiation is 
found to be compatible only with  
evolutionary universes (expanding or contracting ones). For sure, these solutions indicate that the problem is more complex than 
admitted.  

Inspired by the Thermodynamics of Black Holes, Penrose put forward the suggestion that the gravitational field should have an 
associated entropy which, in turn, should be related with an invariant combination involving the 
Weyl tensor \cite{Penrose1}. Remarkably this suggestion allows for a consistent 
set up for cosmology of the Generalized Second Principle of Thermodynamics, which arises in black hole physics, and states that 
the Second Principle should apply to the sum of the entropy of matter with the one of the black hole \cite{Penrose2,HB}. 
The main point of the proposal is that it circumvents the paradox 
of an universe whose initial state is a singularity or a black hole protected by a horizon, and hence with an initial entropy that 
exceeds by many decades of magnitude the entropy of the observed universe. Being highly homogeneous and isotropic, the initial state 
of the universe has necessarily a low entropy\footnote{The low entropy of the highly ``excited'' and hot initial state 
was suggested to be analogous of systems with a negative temperature \cite{Bertolami85}.} 
as the Weyl tensor vanishes for homogeneous and isotropic geometries . The gravitational entropy
will then increase as the Weyl tensor increases as the universe grows lumpier. 

The growth of the total entropy can presumably account for the asymmetry of psychological time as in this way the branching 
of states and outcomes will occur into the future.   

Let us close this subsection with some remarks on some recent ideas developed in the context of superstring/M-theory.  
These suggest a multiverse approach
of the ``landscape'' of vacua of the theory (see e.g. \cite{Bertolami06a} and references therein), that is  
the googleplexus of about $10^{500}$ vacua \cite{BoussoPol}, which are regarded as distinct universes, which asks for 
a selection criteria for the vacuum of our universe. Anthropic arguments \cite{Susskind} 
and quantum cosmological considerations \cite{HMersini} have 
been advanced for this vacuum selection as a meta-theory of initial conditions. These proposals are not 
consensual, but can be seen as a relevant contribution to a deeper understanding of the problem. Of course, 
one should keep in mind that non-perturbative aspects of string theory are still poorly known \cite{Polchinski}.
The multiverse perspective hints at the possibility that different universes may actually interact \cite{Bertolami07}. 
It is suggested that this interaction is regulated by a Curvature Principle and shown, in the context of a simplified model 
of two interacting universes, 
that the cosmological constant of one of the universes is driven toward a vanishingly small value. The core of this 
proposal is to set an action principle for the interaction using the 
curvature invariant $I_{i} = R_{\mu \nu \lambda \sigma}^{i}  R^{\mu \nu \lambda \sigma}_{i}$, 
where $ R_{\mu \nu \lambda \sigma}^{i}$ is the Riemann tensor of each universe. 
The suggested Curvature Principle also hints at 
a solution for the entropy paradox of the initial state of the universe \cite{Bertolami07}. 
For this, one considers the point of view of another universe, from which our universe 
can be perceived as if all its mass were concentrated in some point and hence $I=48 M^2 r^{-6}$, where $r$ is the universe 
horizon's radius and $M$ its mass - using units where $G=\hbar=c=1$. Hence, if the entropy scales with the volume, 
then $S \sim r^3 \sim I^{-1/2}$; if the entropy scales according to the holographic principle, suitable for AdS spaces 
\cite{Fischler,Bousso}, then 
$S \sim r^2 \sim I^{-1/3}$. In either case, given that $I \sim \Lambda^2$ for the ground state, 
one obtains that $S \rightarrow 0$ in the early universe and, $S \rightarrow \infty$ 
when $\Lambda \rightarrow 0$. The latter corresponds to the universe at late time,  
which is consistent with the Generalized Second Principle of Thermodynamics. 

Of course, a multiverse perspective, if taken to its most extreme versions, can lead to intricate problems concerning the 
relationship among the cosmic time of each universe and the ``meta-time'' of the whole network of universes. Only the future will 
tell us whether developments in this direction will be needed to further understand the physics of our universe.

\subsection{Time in quantum gravity}

Quantum gravity, the theory that presumably describes the behaviour of space-time at distances of the order of the Planck 
length is still largely unknown. The most developed programme to understand quantum gravity, superstring/M-theory, 
leads to a quite rich lore of ideas and concepts, but has not provided so far a satisfactory answer concerning 
for instance, the fundamental problem of smallness of the cosmological constant \cite{Witten00}, and exhibits the 
vacuum selection problems discussed above, which seriously threaten the predictability power of the whole approach. 

In order to understand the conceptual difficulties of the quantum gravity 
problem, let us see that from its very beginning, the quantization of gravity poses 
outstanding challenges to the well known and well tested methods of quantum field theory. Indeed, if one considers the metric, 
$g_{\mu \nu}(\vec{r}, t)$, a bosonic spin-two field and attempts its quantization through an equal-time commutation relation for the 
corresponding operator:
\begin{equation}
[\hat{g}_{\mu \nu} (\vec{r},t), \hat{g}_{\mu \nu}(\vec{r'},t)] = 0~, 
\end{equation}  
for $\vec{r-r'}$ space-like, then one faces an indefinite problem: i) In fact, in order to establish that $\vec{r-r'}$ is space-like,  
one must specify the metric; ii) Being an operator relationship, it must hold for any state of the metric; iii) Without 
specifying the metric, causality is ill-defined.

These difficulties compel one to consider a canonical quantization programme based on Hamiltonian formalism 
(see e.g. \cite{BMourao91} and references therein). In this context, one splits space and time and selects foliations 
of space-time where the physical degrees of freedom of the metric are the space-like ones, $h_{ab}=^{(3)}\!\!g_{ab}$. The resulting 
Hamiltonian is a sum of constraints, one associated with invariance under time reparametrization, the others related 
with invariance under 3-dimensional diffeomorphisms. If one considers only Lorentzian geometries (a quite restrictive condition !), 
then only the first constraint is relevant. The solution of the classical constraint is given by:
\beq
H_0=0~,
\eeq
where 
\beq
H_0=\sqrt{h}\left[h^{-1} \Pi_{ab} \Pi^{ab} - ^{(3)}\!\!R\right]~,   
\eeq
$h$ being the determinant of the 3-metric $h_{ab}$, $\Pi_{ab}$ the respective canonical conjugate momentum and $^{(3)}\!\!R$ the 3-curvature.
Quantization follows by turning the momenta into operators for some operator ordering and applying the resulting Hamiltonian 
operator into a wave function, the wave function of the universe, $\Psi[h_{ab}]$:
\beq
\hat{H}_0 \Psi[h_{ab}] = 0~.   
\label{WDW}
\eeq
This is the well known Wheeler-DeWitt equation. 

In this context, the problem of time (see Ref. \cite{Isham} for a detailed account) consists in not having a Schr\"odinger-type equation 
for the evolution of states, but instead, the constrained problem (\ref{WDW}), where time is one of the variables within $H_0$.
Of course, this does not mean that there is no evolution, but rather  
that there is no straightforward way of extracting a variable from the formalism 
that resembles the cosmic time one is used to in classical cosmology. 

Solutions, although partial, include the semi-classical approach \cite{DeWitt,Vilenkin}, where time is identified with the 
scale factor or some function of it, once the metric starts behaving like a classical variable and the wave function of 
the universe admits a WKB approximation. In this instance, the Wheeler-DeWitt equation can be written, at least in the minisuperspace 
approximation, as the Hamilton-Jacobi equation for the action of the WKB approximation. Physically it implies that time is meaningful 
only after the metric becomes classical. 

Another interesting idea is the so-called ``Heraclitean time proposal'' \cite{UW,Bertolami95}. This is based on a suggestion 
by Einstein \cite{Einstein} according to which the 
determinant of the metric might not be a dynamical quantity. In this theory, usually referred to as unimodular gravity, the 
cosmological constant arises as an integration constant and an ``Heraclitean'' time can be introduced as the classical Hamiltonian 
constraint assumes the form \cite{UW}:
\beq
H = \Lambda h^{1/2}~,
\eeq   
and thus, for a given space-like hypersurface $\Sigma$, one can write 
\beq
i{\partial \Psi \over \partial t}= \int_{\Sigma} d^3 x h^{-1/2} \hat{H}_0 \Psi = \hat{H} \Psi~,
\eeq
which has a  Schr\"odinger-like form.

For sure, the problem of time in quantum gravity still remains an open question and the above approaches were presented only 
to exemplify some possible directions for future research.

\subsection{Closed time-like curves and time travel}

As already mentioned, closed time-like curves arise as solutions of Einstein's field equations. These solutions 
include traversable wormholes \cite{Lobo,Visser,Morris,Lobo1}, warp drives \cite{Alcubierre,VisserLobo} and the Krasnikov 
tube \cite{Krasnikov}. One can argue that they are unphysical as they violate the energy conditions \cite{Deser}. 
These solutions correspond to putative forms of time travel and most often bring 
a host of paradoxes of the ancestor's murder type. However, given that the murder of an ancestor 
by a time traveler should be logically inconsistent, 
one could ask whether there should exist global self-consistent conditions to exclude closed time-like curves. These conditions are 
referred to as {\it consistency constraints}. The most discussed of these consistency constraints are the Principle 
of Self-Consistency \cite{Earman} and the Chronology Protection Conjecture \cite{Hawking92}. 

The Principle of Self-Consistency states that events along a closed time-like curve are self-consistent, that is they influence 
each other, but in a self-consistent fashion. Of course, along a closed time-like curve the notion of past or future is ambiguous 
and the causal structure of usual space-times is meaningless. The self-consistent condition establishes that events in the future can 
influence events in the past, but cannot alter them. Hawking's Chronology Protection Conjecture is based on the experimental 
evidence that ``we have not been invaded by hordes of tourists from the future'' \cite{Hawking92} from which it is then 
conjectured that the renormalized stress-energy tensor 
quantum expectation values diverge as they approach closed time-like curves. This divergent behaviour destroys 
the wormhole's structure before the Planck scale is attained. So far, no proof of this conjecture is available.    
 
Thus, one sees that the reality of closed time-like curves may be contested on physical as well as on logical grounds. Nevertheless, 
these solutions are vivid examples of the wealth of structurally distinct solutions of general relativity and show 
how some classes of solutions may require a specific set of criteria to establish their physical reality.

\subsection{A cyclic time ?}

The general theory of relativity allows a for a global dynamical description of the physical space-time and 
for a relation with the history
and evolution of the universe.
The mathematical description of space-time admits a wide range of scenarios, which includes solutions with cyclic nature. 
Already in 1922 Alexander Friedmann (1888 - 1925), the first to study evolving cosmological solutions within general relativity, 
realized that cyclic scenarios existed among his solutions. These involved an expanding universe followed by a recollapse so that 
the universe's radius would eventually vanish from which a new expansion would ensue. Of course, strictly speaking these cycles are not 
mathematically admissible as they are disjoint by a singularity. In 1931, Richard Tolman (1881 - 1948) \cite{Tolman} 
showed that such discontinuity 
was unavoidable at the beginning and at the end of any isotropic and homogeneous closed geometries for a physically realistic 
energy-momentum tensor. Subsequently, he argued that the problem was actually due to the highly symmetric nature of the studied 
solutions and that in a physically realistic universe the discontinuity could very well disappear \cite{Tolman1}. 

A cyclic or ``phoenix'' universe was regarded with sympathy by Einstein and George Gamow (1904 - 1968), who even coined the 
term ``big squeeze'' to denote the final state of collapse - nowadays the term ``big crunch'' is more used. Of course the issue 
of space-time singularities was not fully appreciated then; however, in the 
1960s it was understood, through the Hawking-Penrose singularity 
theorems, that the conditions and the generality of the difficulty could not be overlooked and cosmologists had to accept the reality 
of the space-time singularities. Some relativists argued however, 
that quantum effects could play a role in the process of ``bouncing'' at 
very high densities completing in this way the cycle of a closed universe. John Wheeler, for instance, advocated that 
in the ``bounce'' physical constants would be recycled \cite{Wheeler}.     

More recently, developments in string theory and the related dynamics of branes do open the possibility of reviving the idea of 
a cyclic universe. In the so-called ``ekpirotic'' model \cite{Ekpirotic}, one assumes, as a starting point, the existence of 
two 3-dimensional parallel branes embedded in a higher dimensional space. Our universe corresponds to one of these branes. 
Quantum fluctuations in the other brane would lead to the creation of a third brane, which would be attracted to ours. The ensuing 
impact of the third brane into ours would trigger a release of energy, {\it ekpirosis} in Greek, giving origin to a 
proto-universe, whose subsequent expansion would have properties similar to the ones of a universe just emerging from the inflationary 
process. Thus, this collision process is quite similar to the Big-Bang itself. Of course, whether a universe emerging from the 
ekpirotic process fully resembles our universe, or whether it advantageously replaces the inflationary dynamics, whose most
generic features are consistent with the latest observations of the cosmic microwave background radiation \cite{WMAP3}, is still 
a quite open question. It is interesting that these two competing models have 
a distinct behaviour in what concerns the production of primordial 
gravitational waves. The ekpirotic process tends not to produce too much gravitational waves, while some 
models of inflation do produce a considerably greater amount of gravitational radiation. The possibility of verifying 
the prediction of these models through the observation of gravitational waves is of course quite exciting. 

The cyclic nature of the  ``ekpirotic'' model arises from the fact that after several decades of thousands of millions of 
years after the 
brane collision, our universe will expand to the 
point where stars and galaxies will be all gone and there will be no radiation left. 
This void and cold brane will be very similar to the one before the Big-Bang. Conditions will then be favourable for the creation 
of a third brane from the other original brane and the whole process then repeats itself\footnote{The 
similarity with the Indian mythology 
is compelling. Each cycle is analogous to the ``day of Brahma''. The whole process resembles the ``life of Brahma''.}.

\section{Conclusions and Outlook}
\label{sec:5}

The unification of space and time proposed by Minkowski a century ago allowed for an elegant formulation of the special 
theory of relativity, and was the culmination of more than two thousand years of philosophical and physical research 
on the nature of space and time. The space-time continuum is a basic foundational concept in physics, from elementary 
particle physics to cosmology. The vector space 
structure of the space-time continuum made the transition to the general theory of relativity smooth and quite logical 
once it was understood that, at cosmological scales, the space-time continuum was not an {\it a priori} concept, independent 
of the physical conditions. Furthermore, when analyzing the inner makings of matter, Minkowski's space-time formulation, 
together with quantum mechanics, made possible, through renormalizable quantum field theory, to stretch our knowledge down to 
scales of about $10^{-18}$ m. The research on the matching of general relativity with the quantum nature of matter is still 
in its infancy; however, we already understand that reconciling these two pillars of the XX's century physics 
will require a whole new 
set of ideas, as it may happen, that we may 
have to give up concepts that were supposed to be the starting point of all the modeling of the 
universe, such as that space-time is a continuum and that the fundamental building blocks of matter are not point-like particles.    

These assumptions lead to quite new realms for research and experimentation. They also pose us new technical and conceptual 
problems. These imply that the very principles upon which our theories of space-time were built so far, such as Lorentz invariance, 
CPT-symmetry, the commutative nature of the fundamental dynamical variables and so on, will have to be continuously scrutinised. Their 
breakdown may provide important insights into the nature of the new theories of space-time, matter and the universe.    
Of course, these new theories will have to match smoothly our current physical theories and explain the conditions for the emergence 
of the Minkowski space-time continuum as well as to set the boundaries of validity of general relativity and the emergence of the 
classical features of gravity. The new theories will, like in the case of 
general relativity, pose questions of ontological nature and should 
set criteria for selecting,  among the mathematically consistent solutions, the physical ones which have predictive 
power to explain our world. The most recent developments in the context of superstring/M-theory, the most studied quantum 
gravity programme, suggest that a 
multitude of universes is needed to explain the physics of our universe. This is a somewhat disappointing outcome for a theory that 
naturally unifies quantum mechanics and general relativity. However, this may only reflect the provisional state of our knowledge.  

On the other hand, 
as we have discussed, the quest for the understanding of the ultimate nature of space-time, and the rather special role 
played by time in macrophysics and its various arrows is still largely unknown. If all arrows 
of time can be related with the expansion of the universe, or to 
some new curvature principle that properly accounts for the entropy of the gravitational field, a remarkable new unification 
could be achieved. 

In any case, the quest for the ultimate theories about the nature of space and time have mesmerized human thought for more than 
two thousand years. Till recently, the most insightful 
ideas sprang from philosophical speculation, however since the pioneering work of 
Einstein on relativity and the space-time unification proposed by Minkowski a century ago, physicists have taken the lead in this  
search. A hundred years after the proposal of Minkowski, mankind is about to embark on new expeditions to conquer new continents of 
knowledge through new scientific challenges which include the Large Hadron Collider to search for the nature of mass and new 
symmetries, and new space missions to study the polarization of the cosmic microwave background radiation and to directly 
detect gravitational waves. It is the hope of the whole scientific community that the outcome of these experiments 
will bring precious hints for the understanding of our universe.

%
%
%



\printindex
\end{document}